\documentclass[preprint]{aastex6}

\begin{document}                                                                     

\title{ V659 Cen: System Members Updated 
} 


\author{Nancy Remage Evans}
\affil{Smithsonian Astrophysical Observatory,
MS 4, 60 Garden St., Cambridge, MA 02138; nevans@cfa.harvard.edu}

\author{Charles Proffitt}
\affiliation{Space Telescope Science Institute, 3700 San Martin Drive, Baltimore, MD 21218}

\author{Pierre Kervella}
 \affil{LIRA, Observatoire de Paris, Universit\'e PSL, Sorbonne Universit\'e, Universit\'e Paris Cit\'e, CY Cergy Paris Universit\'e, CNRS, 5 place Jules Janssen, 92195 Meudon, France.
    \and 
    French-Chilean Laboratory for Astronomy, IRL 3386, CNRS and U. de Chile, Casilla 36-D, Santiago, Chile}

\author{Joanna Kuraszkiewicz}
\affil{Smithsonian Astrophysical Observatory,
MS 67, 60 Garden St., Cambridge, MA 02138; jkuraszkiewicz@cfa.harvard.edu}

\author{H. Moritz G\"unther}
\affil{Massachusetts Institute of Technology, Kavli Institute for Astrophysics and
Space Research, 77 Massachusetts Ave, NE83-569, Cambridge MA 02139, USA}

\author{Richard I. Anderson}
 \affil{Institute of Physics, \'Ecole Polytechnique F\'ed\'erale de Lausanne (EPFL), Observatoire de Sauverny, Chemin Pegasi 51B, 1290 Versoix, Switzerland}

\author{Alexandre Gallenne}
\affiliation{Instituto de Alta Investigaci\'on, Universidad de Tarapac\'a, Casilla 7D, Arica, Chile}

\author{Antoine M\'erand}
 \affil{European Southern Observatory, Karl-Schwarzschild-Str. 2, 85748 Garching, Germany}

\author{Boris Trahin}
\affil{Space Telescope Science Institute, 3700 San Martin Drive, Baltimore, MD 21218}

\author{Giordano Viviani}   
\affil{Institute of Physics, \'Ecole Polytechnique F\'ed\'erale de Lausanne (EPFL), Observatoire de Sauverny, Chemin Pegasi 51B, 1290 Versoix, Switzerland}

\author{Shreeya Shetye}   
\affil{Instituut
voor Sterrenkunde, KU Leuven, Celestijnenlaan 200D bus 2401, Leuven, 3001,
Belgium}


 \begin{abstract} V659 Cen is a classical  Cepheid which is  part of a multiple system. Previous observations have shown that a hot companion  dominates an ultraviolet spectrum and  a cooler main sequence star dominates an XMM-Newton   spectrum. The {\it Hubble Space Telescope (HST)} Space Telescope Imaging Spectrograph (STIS) spectra discussed here    spatially resolve the components and show that the secondary in the spectroscopic binary 
   with the Cepheid is the low mass star, and  the hottest star in the system is the  outer
   companion.  In addition a fourth star is a likely member of the system based on {\it Gaia}
   data. A new orbit is derived which includes new radial velocities. 
\end{abstract}




\section{Introduction}





For massive and intermediate mass stars it has become increasingly clear that
they are typically found in binary or multiple systems.  Happily there are an
increasing number of observational techniques to determine the structure of the
systems, especially through the use of different wavelength regions.  V659 Cen
provides a case study of the  analysis of such a system.

In 1959, Abt stressed the surprise that only 2\% of Cepheids were found in
binary systems, as compared with 23 \% of O and B stars which develop into
Cepheids.  Since then a much higher fraction has been found from studies sensitive to hot (high mass) companions (Evans 1992a), low mass companions (Evans, et al. 2022), and proper motions (Kervella, et al. 2019a) as well as extensive new velocity data (Anderson et al. 2024; Sheyte et al. 2024).  A comprehensive summary of multiplicity properties requires summarizing results from a range of companion masses as well as a range of separations.  As an example, Evans, et al. (2015) found a preferred binary fraction of 29\% $\pm$ 8\% from a sample of stars with radial velocities covering 20 years.  An additional challenge is that many systems with an intermediate mass primary have more than two components, which increases the complexity of interpreting the results.   

 The V659 Cen system is a case in point.
The V659 Cen system contains up to four members.  Several wavelength regions and
techniques have been used to gradually identify the properties of the components. 
 While previous studies have determined some properties of the members, they have not determined how the system is structured in terms of (for instance) whether the hottest star is a member of the spectroscopic binary with the Cepheid, or a more distant companion. The new data in this report addresses this question. This information is important, for instance, in assessing whether the system can be used in future work such as determining masses. 







\section{The V659 Cen System}

Among the extensive list of studies to investigate Abt's finding  was that of
Lloyd Evans (1982), using a list of southern Cepheids.  V659 Cen was one he
considered, however he found only $\pm$ 2 km s$^{-1}$ between observations from
different years, and concluded that binarity was dubious.  Photometry was scarce and
did not allow tests for a companion.
The detection of a hot star in an {\it International Ultraviolet Explorer (IUE)} satellite
spectrum of V659 Cen made the binary nature of the Cepheid clear (Evans 1992b).

Although V659 Cen was clearly identified as a member of a multiple star system, insights
into the components have continued from a combination of instruments.
 This section summarizes what is known about the properties of the components.  A schematic summary of the system is provided in the final section of this paper, updated from a figure in Evans, et al. (2022)

\subsection{ The Resolved Companion}
A  companion was resolved in  {\it Hubble Space Telescope (HST)} Wide Field Camera 3
 (WFC3)
image (Evans, et al. 2013).  The observation was part of a snapshot survey of Cepheids
made in two filters F621M and F845M. The companion is  close to
 the much brighter Cepheid and lies on a diffraction spike, so accurate photometry
was not immediately possible from that image.
However further processing to remove the  Cepheid (Evans, et al.
2020) provided an approximate separation of 0.6$\arcsec$  between the Cepheid and the companion.
Using a distance of 1089 pc
from Section ~\ref{ebv.sed}  this becomes 653 au.


\subsection{The Low-Mass Companion}

The V659 Cen system was observed with the X-ray satellite {\it XMM-Newton}  (Evans, et al. (2016)  to test several possible low mass
resolved companions found  in the HST survey. X-ray sources would have  been
expected for F, G, or K main sequence stars as young as Cepheids.
None of the suggested resolved
companions  coincided with X-ray sources.    However,
an X-ray source was found  coincident with the Cepheid (Evans, et al. 2020).  This
was discussed further by Evans, et al. (2022) in comparison with what is known
about X-rays produced by Cepheids.  The source has an X-ray luminosity
log L$_X$ = 29.51 ergs s$^{-1}$, which is much brighter than anything seen in
Cepheids, particularly at the phase  (0.14) of observation  where Cepheids do not show X-ray brightening.
However, it is in
 agreement with the luminosity expected for a young cool main sequence
companion.  


\subsection{{\it Gaia} Companion}

  Based on {\it Gaia} DR2  Kervella, et al. (2019b)  found an additional wide
M3 V companion  separated by 62$\arcsec$ based on similar parallax, and small differential
tangential velocities.  Based on the differential tangential velocities it is
considered gravitationally bound.  This component with an M spectral type  would not be expected to be detected in the X-ray observation.

\subsection{Orbit}
Based largely on velocities from Anderson, et al (2024), Shetye, et al (2024)
derived an orbit for  V659 Cen.  However, they described it as tentative because
the data allow  orbital periods of either 6000 or 9000 days.





\subsection{Pulsation mode}
V659 Cen has a low pulsation amplitude, which is suggestive
(though not definitive) of pulsation in the first overtone rather than the
fundamental mode.  The pulsation period itself (5.621800   $^d$) is longer than is
typical for first overtone pulsators.  The 
 VELOCE study of Cepheid velocities (Anderson et al. 2024) provides data for
Fourier coefficients.  The extensive experience of Hocd\'e, Moskalik et al. (2024)
in mode identification emphasized the use of Fourier phases.  
V659 Cen is firmly in the range of overtone pulsators in their plot of the Fourier phase parameter  $\phi_{21}$ as a function of log P (pulsation
period).  While
overtone pulsators with a pulsation period this long are rare, others are known. We
conclude that V659 Cen is pulsating in the first overtone.

\section{STIS Observations}
\subsection{Companion identification}
A remaining puzzle for the V659 Cen system is whether the hottest star in the
system which dominates the {\it IUE} spectrum is the companion in the
spectroscopic binary or the resolved  0.6$\arcsec$  companion.

\subsubsection{Observation Plan}
In order to answer this question, we obtained {\it Hubble Space Telescope  (HST)}
Space Telescope Imaging Spectrograph (STIS) G140L spectra.
Details of the observing sequence are provided in
Table~\ref{exp}.
Full details of STIS parameters are provided in the
STIS Instrument Handbook
\footnote{https://www.stsci.edu/hst/instrumentation/stis}.
However aspects of the observations which affect our
interpretation are discussed here.
As our estimated brightness for this star  approaches the allowed
STIS bright object limit for G140L, we were limited to using either
the very narrow 52X0.05 aperture or one of the neutral density filters.
We chose to use the 52X0.05 aperture at a position angle intended to
align the two stars along the slit, as well as a slitless exposure
with the F25NDQ1 aperture to ensure we obtained at least one exposure
that included an unvignetted spectrum of any UV bright object near
V659 Cen.  The orientation adopted was based on a preliminary analysis
of the WFC3 images; however, since both components are saturated in
those images, and the companion falls on one of the diffraction spikes
from the brighter Cepheid, the position angle was uncertain.
In addition, we
could not rule out that the PA of the companion might have changed
since the WFC3 images due to either orbital motion or differential proper motion.
    
Our observations were obtained on July 8, 2020. A target acquisition (TA) was
done, followed by a peakup in the 52X0.05 aperture using the G230LB grating.
A 1500s spectroscopic exposure was then taken in the F25NDQ1 aperture
with the G140L grating, followed by a 300s G140L exposure in the 52x0.05
aperture.  A G140L wavecal exposure was then taken using the 52X0.05 aperture.
    
\begin{deluxetable}{lrrrr}
\tablecaption {STIS exposures of V659 Cen\label{exp}}
\tablewidth{0pt}
\tablehead{ 
  \colhead{Dataset} &  \colhead{ Optical }  & \colhead{Aperture}  & \colhead{Exposure Time} & \colhead{OBSMODE}  \\
  \colhead{} &  \colhead{Element }  & \colhead{}  & \colhead{s}  & \colhead{} \\
}
\startdata
oe2e10pyq &	MIRVIS &	F28X50OII &	2 &	ACQ \\
oe2e10pzq &	G230LB &	52X0.05 &	5 & 	ACQ/PEAK \\
oe2e10010 &	G140L & 	F25NDQ1 &	1500 &	TIME-TAG \\
oe2e10020 &	G140L &	52X0.05	& 330  &	ACCUM \\
oe2e10q5q &	G140L &	52X0.05 &	27  &	WAVECAL \\
\enddata
\end{deluxetable}

\subsubsection{Acquisition image}

\begin{figure}
  \plotone{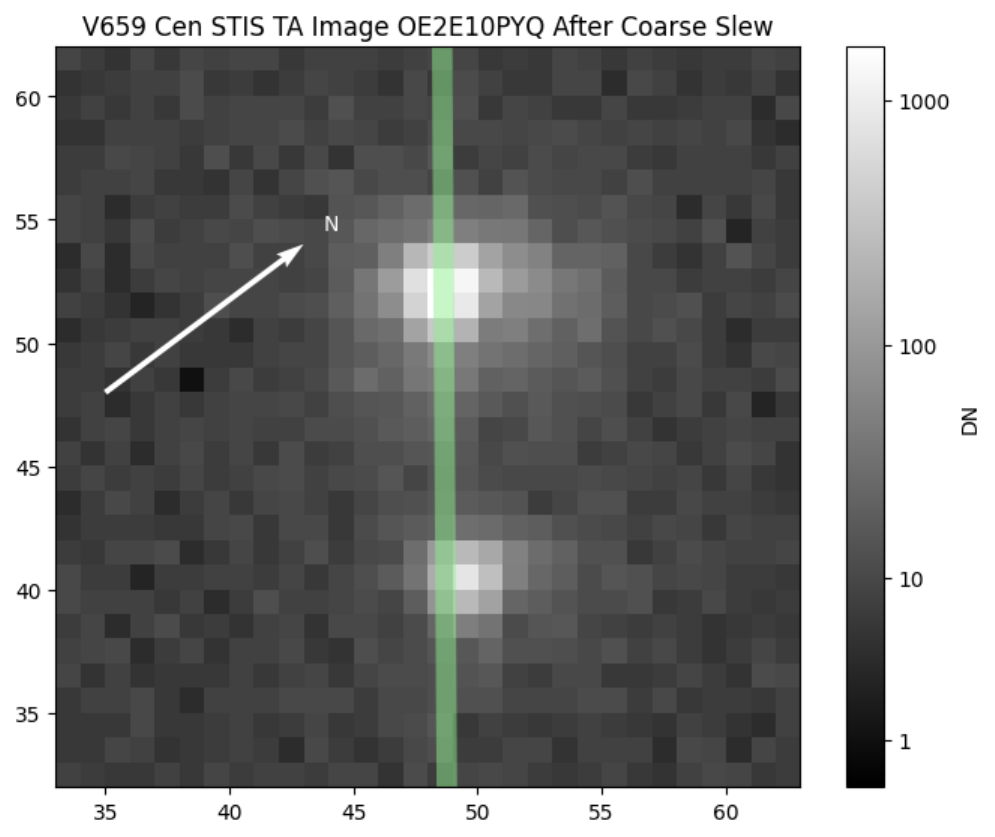}
  \caption{  The second image of V659 Cen that was taken as part of the
    target acquisition exposure OE2E10PYQ. The Cepheid and spectroscopic
    binary companion is the
    brighter upper star; the resolved companion is the lower star.
    We superimpose a vertical line
    in green showing our best estimate of the alignment of the 52X0.05
    slit during the subsequent spectroscopic exposure.
\label{acq}}
\end{figure}

In Fig.~\ref{acq}, we show the STIS target acquisition (TA) image
that was taken using the narrow band F25X50OII filter, which has
an 80 \AA\ FWHM with a central wavelength of 3740 \AA. Both components
are clearly visible. While STIS TA images are not optimized for photometry,
we estimate a flux
$F_\lambda$(3740\AA) $\approx$ 6.35$\times10^{-13}$ ergs cm$^{-2}$s$^{-1}$ \AA$^{-1}$ for the
companion, with the Cepheid being about 2.5X brighter. We superimpose
on the TA image our best estimate of the position of the 52X0.05 slit
relative to the stellar targets. It is apparent that the 0.6" companion
(lower star)
was centered outside the slit with an estimated offset of about 0.045"
from the center of this 0.049" wide aperture. So we  expect that
any spectrum obtained of the companion with the 52X0.05 would just be
catching the wing of the PSF.


The STIS target acquisition takes one image at the initial pointing,
and another after the ``coarse locate phase'' which is intended to move
the target to a fixed location on the detector.  We measured the
location of both the primary and secondary in both images, and found
values for the separation of 0.588 and 0.595 $\arcsec$, and for the position
angle of 237.74 and 237.79 degrees.   This position angle differs
from the result of $\theta = 234.9\,$degrees quoted by Evans et al.
(2020) from the WFC3 images, but it is not clear if this due to the
difficulty in measuring the saturated WFC3 image or if it represents
an actual change in position angle.

 Identifying which star in the system is the hottest star (dominates in the ultraviolet) requires both determining the location of the hottest star and also the flux calibration of the spectrum.  These parts are discussed in the next two sections.

\subsubsection{Identifying the Source of the UV bright Spectrum}

A single bright UV spectrum is visible in each of the G140L spectra.
 While the cross-dispersion location of the spectrum in a
 STIS MAMA  (Multimode Anode Microchannel plate Arrays)
   exposure can vary due to the monthly offsetting of the
grating which is intended to spread out the charge extraction on
the MAMA detector, as well as due to non-repeatability of the grating
mechanism itself, any offset can be measured from the wavecal exposure
taken adjacent to the science exposures.  We examined a number of
recent STIS G140L observations of bright sources centered in one of
the 52" long apertures, and found that the mean value for
A2CENTER - SHIFTA2 = 387.9 pixels  with a scatter of about 1 pixel,
where A2CENTER is the spectrum location found by the extraction
routine, and SHIFTA2 is the cross-dispersion offset found by the
wavecal.  For our 52X0.05 exposure we instead find
A2CENTER - SHIFTA2 = 362.66, which is about 0.63" below the
expected location. Thus, the spectrum found by the extraction
routine is clearly that of the 0.6" companion rather than of
the Cepheid itself. 

\begin{figure}
  \plotone{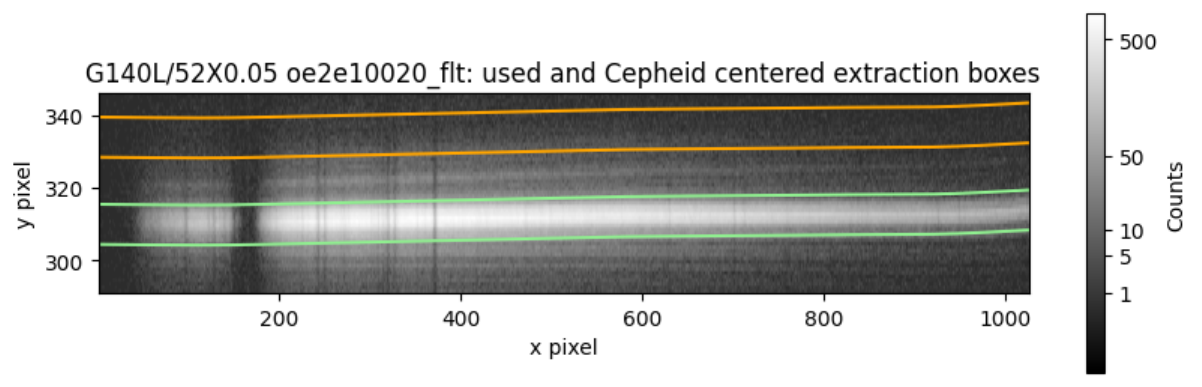}
  \caption{The flat-fielded image of the G140L spectrum of V659 Cen taken
    in the 52X0.05 aperture. The green lines show the upper and lower limits
    of the extraction region used for the 1D spectroscopic extraction. The
    yellow lines show the location the extraction region would have if it
    were instead centered on the Cepheid, but there is no evidence of a
    second star in this image.  The most prominent spectral features
    include multiplets of \ion{C}{3}\ at ~1175\AA, (~col. 95),
    Lyman $\alpha$(~1215\AA, centered near col.\ 165),
    \ion{Si}{2} ~1260, 1265, 1304, and 1309\AA,
    (~cols. 241, 249, 316, 325), and \ion{C}{2}\ 1335\AA\ (~col.  329).
\label{spec_image_52}}
\end{figure}

\begin{figure}
  \plotone{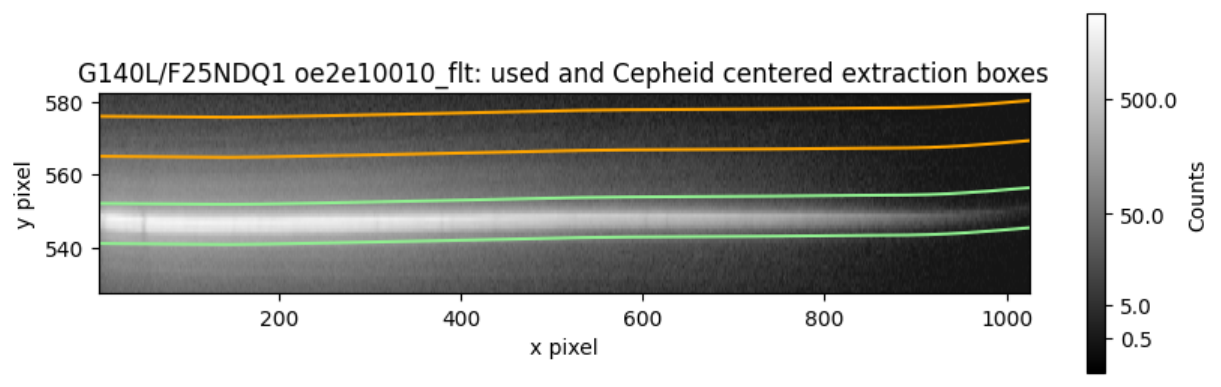}
  \caption{The 2D flat-field image of the G140L exposure taken with the
    F25NDQ1 filter. Conventions are the same as in
    Fig.~\ref{spec_image_52}.  The sharp feature seen near pixel
    50 corresponds to the \ion{C}{3} multiplet near
    1335.
\label{spec_image_f25ndq1}}
\end{figure}

In Figure \ref{spec_image_52}, we show the 2D flat-fielded MAMA
image for the 52X0.05 G140L exposure.  We mark the boundaries
of the extraction region used by CALSTIS, as well as the boundaries
that would have been used for a spectrum centered on the Cepheid
itself.  Figure \ref{spec_image_f25ndq1} shows the slitless spectrum
obtained through the F25NDQ1 filter. Note that the default position
of the target in the F25NDQ1 filter is offset from that of the
52X0.05 aperture by $-7.98$ pixels in the dispersion direction and by $+5.83$
pixels in the cross-dispersion direction.  This shift of the target position
along the dispersion direction changes the wavelength range that falls
on the detector. The 52X0.05 spectrum covers ~ 1120 to 1716 \AA, while
the F25NDQ1 spectrum covers ~ 1310 to 1907 \AA. Neither image shows
any evidence for a second source.

The hottest star is thus Component B in the system as summarized in the final section of the paper. The Cepheid belongs to Component A. 

\subsubsection{Calibration of the G140L spectra}
The F25NDQ filters are one of the more rarely used STIS observational modes.
Indeed, for the combination of G140L + F25NDQ1 there is only one other dataset:
O3ZX080R0, a calibration observation of the white dwarf standard GD153 taken
in 1997. The STIS instrument report (Proffitt STIS ISR 2003-001) that discussed
on-orbit measurement of the MAMA filter throughputs did not include discussion
of the F25NDQ filters, and so some additional uncertainty in the calibration
might be anticipated.

{\bf Wavelength offset in the F25NDQ1}
Given that the position angle used resulted in a displacement of the target
along the dispersion direction from the nominal position for that aperture,
we should expect an offset in the wavelength solution, especially for the
slitless F25NDQ1 exposure. For the 52X0.05, the narrow aperture itself
will limit the size of any offset to $<$  the slit width or less than
1 MAMA pixel, although the exact shift will depend on the distribution
of light in the aperture.  However, we find that to align the absorption
lines in the two spectra, we need to shift the F25NDQ1 exposure by about
5.5 pixels  along the dispersion (X) direction,
which is much larger than the ~ 1.5 to 2 pixels we would have
expected from our analysis of the target acquisition image. One possibility
is that the refraction correction for this filter in the FUV is not properly
accounted for in the calibration files. Unfortunately, the GD153 spectrum has
no spectral features in the wavelength range covered by the G140L with that
filter and so cannot be used to check the quality of the wavelength calibration.

\begin{figure}
  \plotone{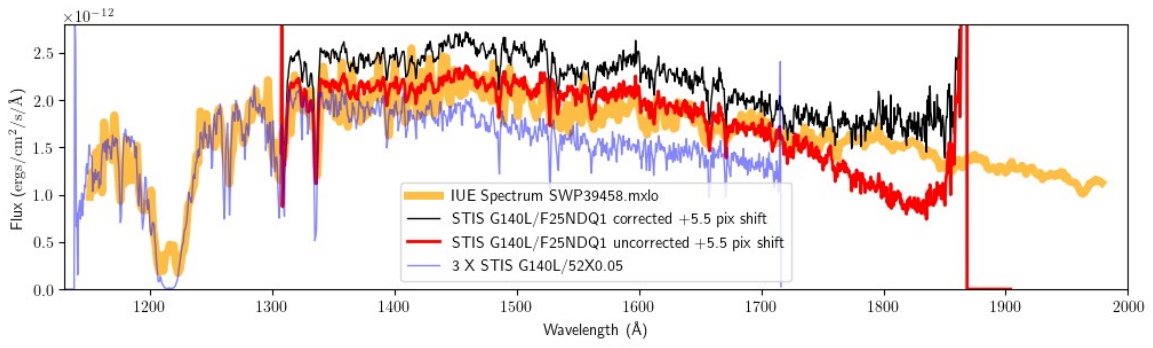}
  \caption{ The STIS and IUE FUV spectra of V659 Cen are compared.
    For the spectrum obtained with the STIS G140L + F25NDQ1,
     the flux both with and without the correction
    implied by the earlier standard star spectrum obtained
    with that filter/grating combination are shown. The flux of 52X0.05
    spectrum has been multiplied by 3. For the F25NDQ1 spectrum
     the 5.5 pixel shift needed to align with the 52X0.05 spectrum
    is included.
\label{specfig}}
\end{figure}

{\bf Flux calibration of the F25NDQ1 spectrum}
The normal wavelength range for STIS G140L spectra is ~ 1120
to 1716 \AA, but the shifted position of the F25NDQ1 aperture
changes the range covered to ~ 1307 to  1904 \AA. For wavelengths
longer than the normal G140L range, the sensitivity function,
as well as the time dependent sensitivity corrections are
apparently extrapolated to cover the additional wavelength
range, and so it is not surprising that the flux calibration
appears to have large errors at those wavelengths. The extracted
spectrum for our target as well that for the GD 153 spectrum shows
qualitatively similar deviations as a function of wavelength in
that range. However, what is somewhat surprising is that the
extracted flux for the GD 153 calibration exposure is also
systematically low by 10 to 15\% between 1307 and 1904 \AA\
when compared to the CALSpec model for that standard star.
The observation of GD153 did use a POS TARG to place the star
at a different location within the aperture, and so it is
possible that there is some spatial variation in the throughput
over the NDQ1 quadrant of the filter. Also, while the STIS aperture
reference file   \texttt{y2r1559to\_apt.fits}
 states that the throughput for
the F25NDQ1 aperture was modified to extrapolate the throughput
to shorter wavelengths and adjusted based on on-orbit imaging data,
there is no mention of having used the available spectroscopic
observations to check or update the throughput as a function of
wavelength for this aperture. While we can correct our extracted
flux from the F25NDQ1 observation for the V659 Cen companion using
a smoothed ratio of the extracted to model flux found for GD 153
in this mode, the resultant flux must be considered to have larger
than usual uncertainties. 
In figure \ref{specfig} we compare the IUE spectrum with the two
STIS G140L spectra. As expected, the 52X0.05 spectrum which was
offset from the target is low by about a factor of 3. The SED
shown by this spectrum is somewhat steeper than either the IUE
or the F25NDQ1 spectrum, but this is likely due to variations
of the PSF width with wavelength affecting the amount of light
that gets into the slit. The F25NDQ1 G140L spectrum without
the flux correction implied by the GD 153 observation is
actually in better agreement with the IUE spectrum below
1710 \AA\ than is the corrected version. We conclude that
the IUE spectrum gives the most reliable measure of the
level and shape of the companion's spectral energy distribution. 
The STIS spectra do have higher S/N and higher spectral resolution
than does the IUE spectrum. The 52X0.05 spectrum also provides a
measure of the Ly$\alpha$ profile with far less geocoronal
contamination that does the IUE observation.  Numerous sharp
features are visible in the STIS spectra. The most prominent
features include Lyman $\alpha$, (~1215\AA), as well as multiplets
of \ion{C}{3}\ near ~1175\AA, \ion{Si}{2} ~1260, 1265, 1304, and
1309\AA, and \ion{C}{2}\ ~1335\AA. Lines of \ion{Si}{3} and
\ion{Si}{4} can also be seen. However, many of these features
will also include contributions from ISM and/or circumstellar
material and this would complicate any attempt to use these
lines to directly constrain the parameters of the companion star.


\section{Orbit}

\begin{figure}
  \plotone{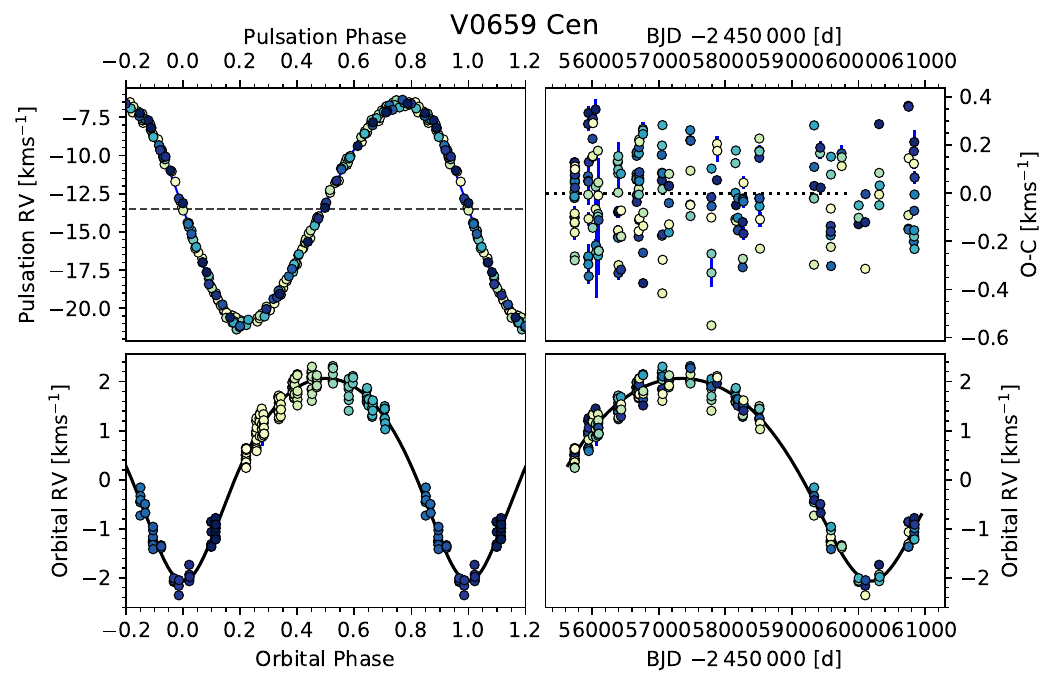}
  \caption{
 Radial velocities of V659 Can. Upper panel left: the pulsational velocity curve as a function of pulsation phase ($P_{\mathrm{puls}} = 5.624$\,d), color-coded by orbital phase ($P_{\mathrm{orb}}=5708$\,d).}  Upper panel right: fit residuals against BJD, color-coded by pulsation phase. Lower panels: the isolated orbital velocity curve as a function of (and color-coded by) orbital phase. Lower panel right: the isolated orbital velocity as a function of BJD color-coded by pulsation phase.
    \label{vr.ria}
\end{figure}

The SB1 nature of V659 Cen was reported by Shetye et al. (2024).  The SB1 system is made up of Components Aa (the Cepheid) and Ab.  This is 
based on high-precision
radial velocity observations collected for the VELOcities of CEpheids project
(VELOCE; Anderson, et al. 2024) using the Swiss 1.2m Euler telescope at ESO's La Silla
Observatory in Chile. VELOCE published available time-series measurements until
5 March 2022. Additional observations carried out as part of the VELOCE project
(for details Anderson et al. 2024) have been collected until 17 June 2025
and successfully resolve the ambiguity on the orbital period pointed out by
Shetye et al. (2024) in favor of the shorter orbital period of $5708 \pm 44$\,d.
These observations will be made publicly available as part of a second VELOCE
data release and are available ahead of time upon reasonable request to
R.I. Anderson. Following the methodology presented in Shetye et al. (2024),
we determined the orbital solution listed in Table~\ref{orb}.
and illustrated in Fig.~\ref{vr.ria}.
The figure shows the full data set and the pulsation and orbital solutions. 
While we consider the orbital solution to be robust at this point, we caution
that the stated uncertainty on the orbital period may not fully reflect that
the baseline of the available data set currently covers $89\%$ of the best-fit
orbital period. Nevertheless, semi-amplitude and eccentricity are well sampled
by the available data.



Revised orbital parameters are listed in Table~\ref{orb}.

\begin{deluxetable}{lr}
\tablecaption{V659 Cen Orbit\label{orb}}
\tablewidth{0pt}
\tablehead{
  \colhead{} &  \colhead{  }   \\
}
\startdata
 V$_\gamma$ (km s$^{-1}$)  & -13.494  $\pm$  0.023  \\
P$_{orb}$ (days)        &   5708 $\pm$   44 \\
e                      &  0.221  $\pm$  0.010  \\
T$_0$  (JD)            & 2460187  $\pm$ 63  \\
$\Omega$ (deg)         & 179.7   $\pm$  3.6 \\
K (km s$^{-1}$)          &  2.067  $\pm$  0.021 \\
a sin i (au)           &  1.058  $\pm$   0.013 \\
f(m)  (M$_\odot$)       &  0.00484 $\pm$  0.00015  \\
\enddata
\end{deluxetable}
 


The a sin i from the orbit can be used to make an initial estimate of the separation of the
stars in the spectroscopic orbit.  Using masses of 5 and 1 M$_\odot$ for the Cepheid and the
companion respectively, 0.5 for sin i and the distance from Section~\ref{ebv.sed},
the separation is estimated to be 0.01$\arcsec$.

\section{Energy Distribution}
For completeness, the temperature of the hottest star  Component B is determined.

\subsection{E(B-V)}\label{ebv.sed}

A valuable addition to the data for V659 Cen since discussion of reddening in 1992
(Evans, 1992b) is the B, V, and I$_C$ photometry  from Berdnikov, et al. (2015)
for the combined system members (Table~\ref{corr}).
 From these colors  the E(B-V) was calculated from the appropriate formula of Fernie (1990) (Table~\ref{ebv})

To derive the effect of the companion on the colors, the Cepheid period was
fundamentalized using the formula of Pilecki (2024) for Milky Way Cepheids to
8.1543$^d$.  Using the PL relation of Cruz Reyes and Anderson (2023) the
M$_V$ = -4.15 mag for the Cepheid. The PL relation is used in preference to 
a {\it Gaia} parallax because of the brightness
and multiplicity of the system.  This corresponds to a distance of 1089 pc.
 Using a preliminary estimate of the temperature of  the companion (16000 K) the temperature corresponds to spectral type B5.3 V based on the temperatures for MK standards (Evans, et al. 2025).  This
spectral type has M$_V$ = -1.11 mag (Drilling and Landolt 2000).  Using this
and the related colors (B-V = -0.15 mag and V-Ic  = -0.16 mag) the magnitudes
of the companion and the Cepheid were computed (Table~\ref{corr}). The revised
E(B-V) is presented in Table~\ref{ebv}.  Since the E(B-V) was changed very little, there was no need to iterate further.

    
\begin{deluxetable}{lrrr}
\tablecaption{Correction for the Companion\label{corr}}
\tablewidth{0pt}
\tablehead{
  \colhead{} &  \colhead{ $<$B$>$ }  & \colhead{$<$V$>$}  & \colhead{$<$I$>$} \\
  \colhead{} &  \colhead{ mag }  & \colhead{mag}  & \colhead{mag} \\
}
\startdata
Cepheid + Companion  & 7.356    &  6.613   &   5.735    \\  
Cepheid               &  7.481    &  6.677    &  5.766    \\
Companion            &  9.767   &   9.717   &   9.629    \\
\enddata
\end{deluxetable}

\begin{deluxetable}{lrrr}
\tablecaption{Reddening\label{ebv}}
\tablewidth{0pt}
\tablehead{
  \colhead{} &  \colhead{  $<$B$>$ - $<$V$>$ }  & \colhead{ $<$V$>$ - $<$I$>$}  & \colhead{E(B-V) } \\
  \colhead{} &  \colhead{ mag }  & \colhead{mag}  & \colhead{mag} \\
}
\startdata
Original   & 0.743       &   0.878              & 0.216      \\
Corrected   &   0.804      &     0.911             &    0.207       \\
\enddata
\end{deluxetable}

\subsection{Companion Temperature}
\begin{figure}
  \plotone{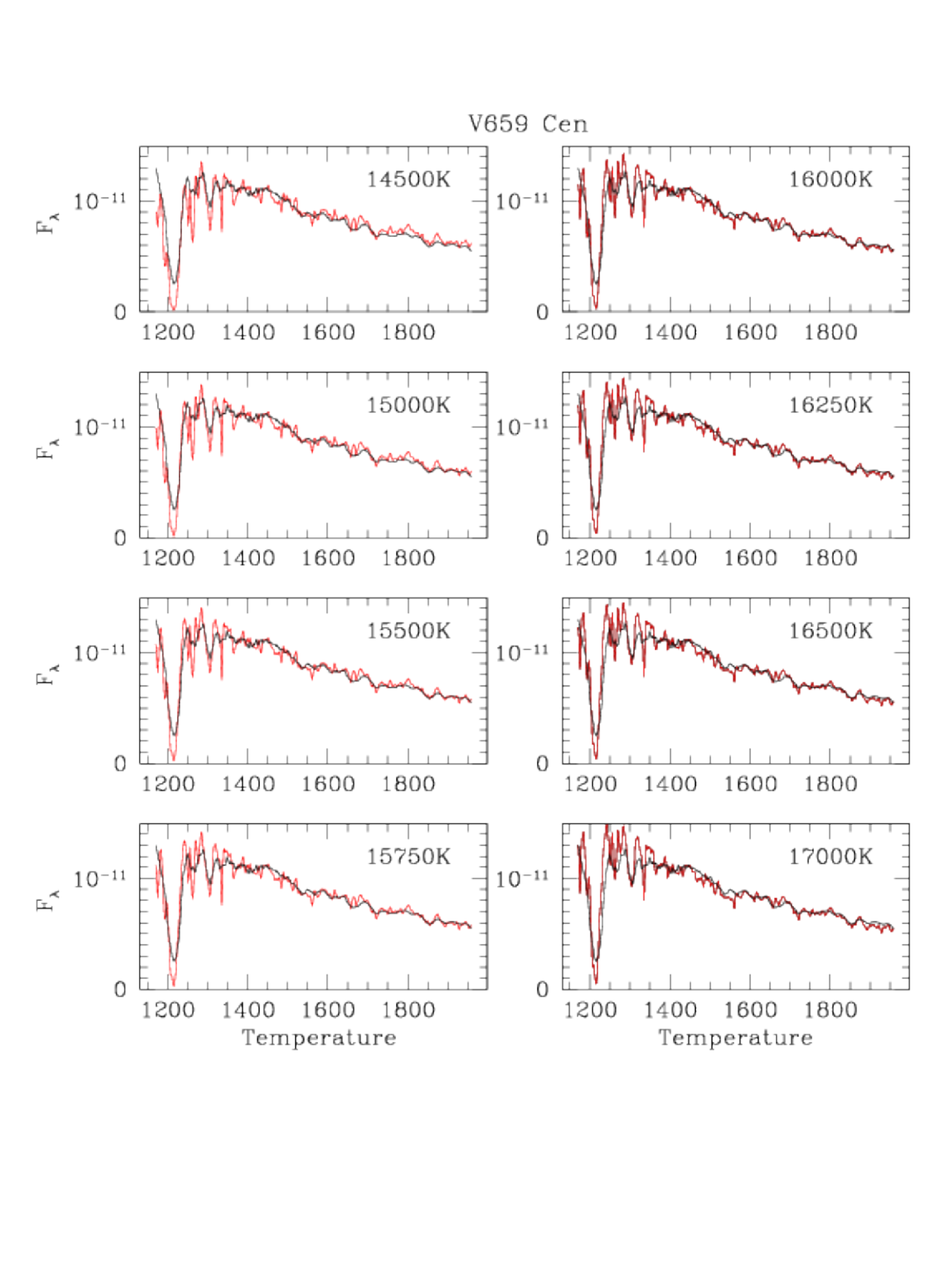}
\caption{Energy distribution comparisons for V659 Cen  Companion B.  Comparisons are for the
    spectrum (black) with model atmospheres (red) for a series of temperatures from
         14500 to 17000 K. The X axis is wavelength in \AA; 
  the  Y axis is flux in ergs cm$^{-2}$ s$^{-1}$ \AA$^{-1}$.
    \label{v659.mod}}
\end{figure}

\begin{figure}
  \plottwo{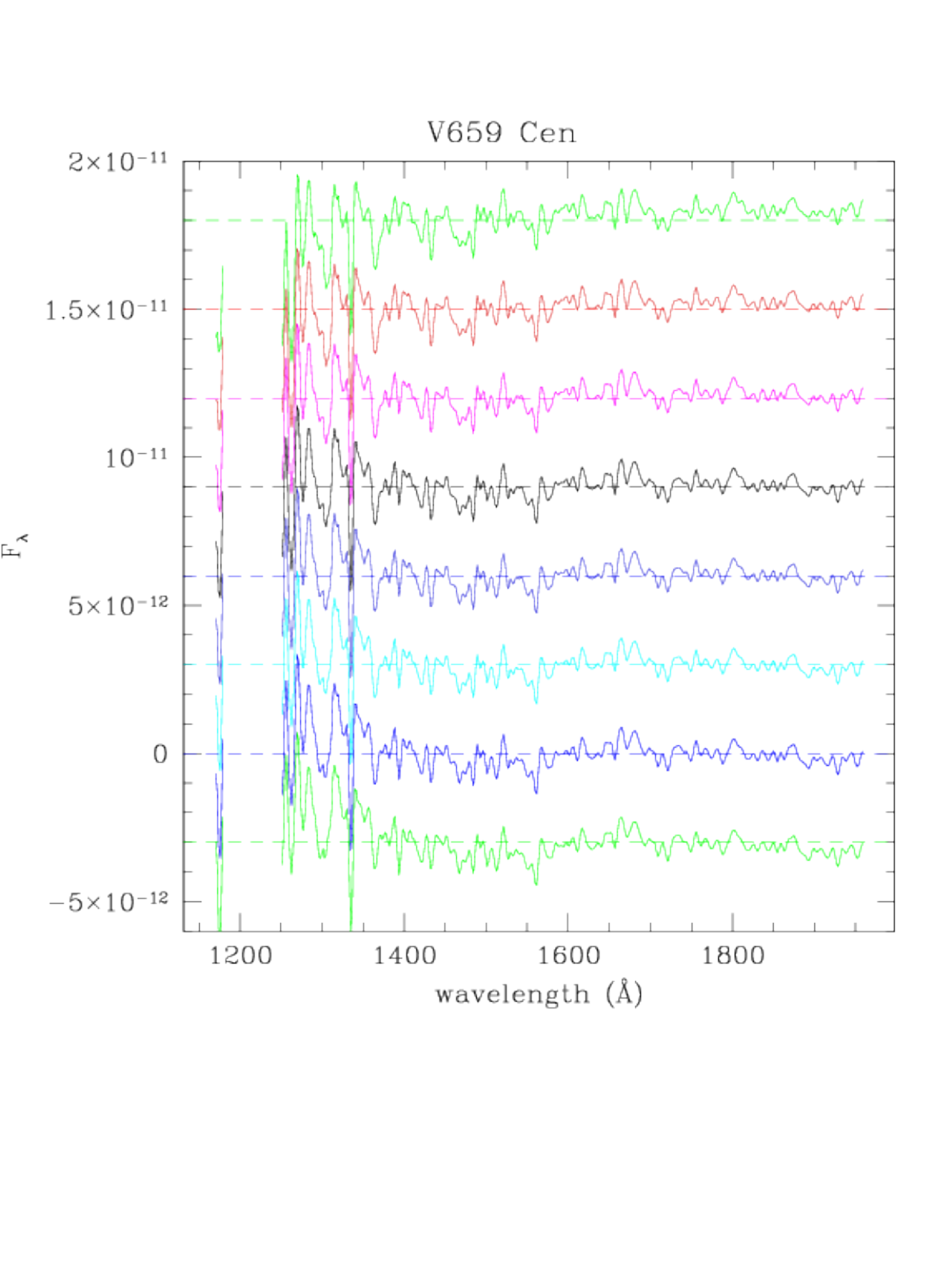}{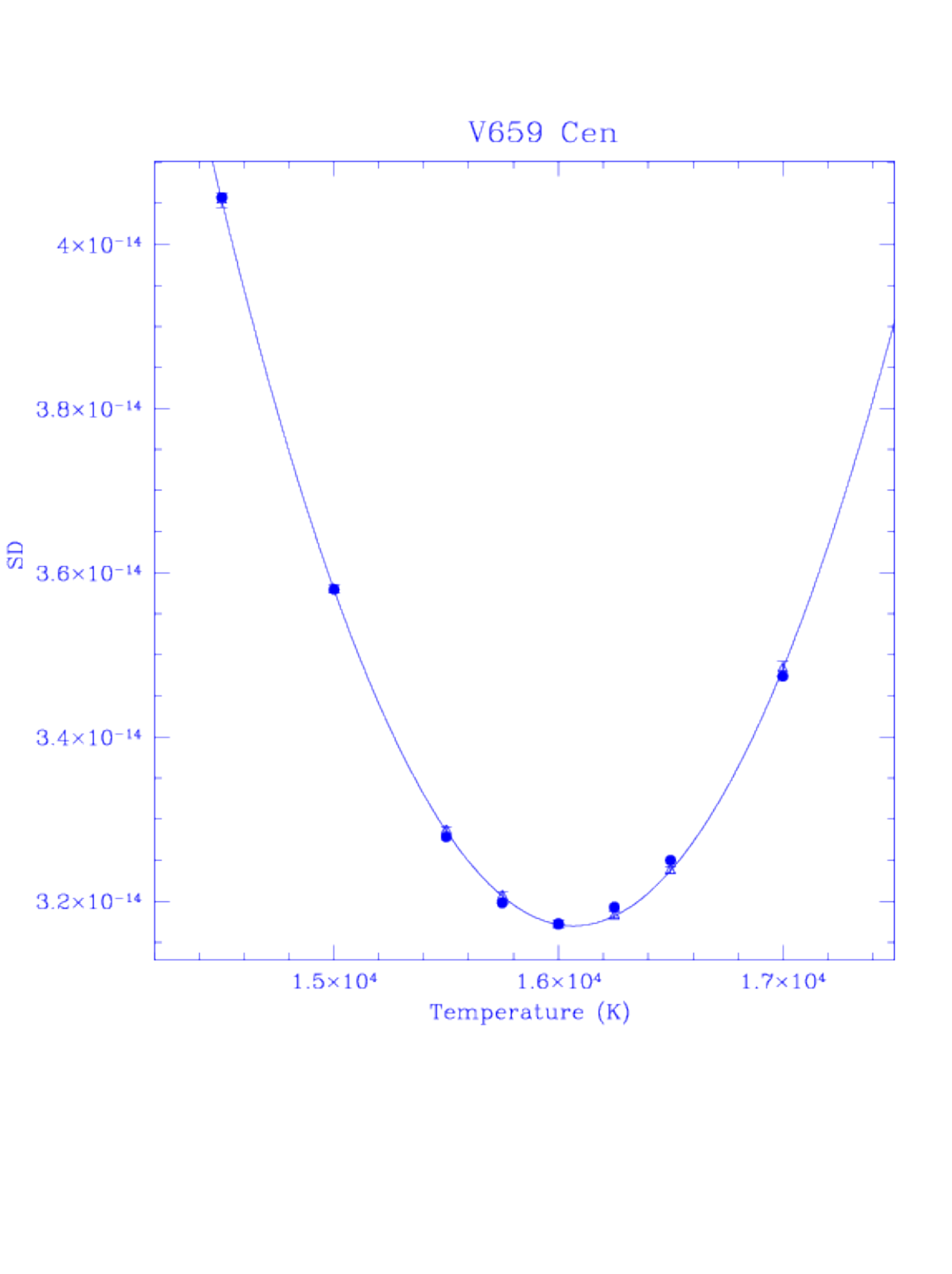}
  \caption{Energy distribution  V659 Cen  Companion B.  Left:  The difference between the
    spectrum and the model.  Model temperatures (starting from the top) are  14500 15000    15500 15750 16000 16250 16500 17000  K. The X axis is wavelength in \AA; 
    the  Y axis is flux difference in ergs cm$^{-2}$ s$^{-1}$ \AA$^{-1}$;
    the models are offset for clarity. 
 Right: The standard deviations from the spectrum-model comparison for
  V659 Cen as the
  temperature of the models is changed.  Dots: the values of standard deviation;
  triangles: the parabola fit.
\label{v659.dif}}
\end{figure}

The  temperature for Companion B was determined from the unreddened IUE spectrum
SWP 39458.  The temperature was found from comparison with
BOSZ atmospheres (Bohlin, et al. 2017)
in the same way as for V350 Sgr (Evans, et al. 2025).  The comparisons
are shown in Fig~\ref{v659.mod}, together with differences from
the models and the parabola fitted to the standard deviations of the
differences in  Fig~\ref{v659.dif}.  The companion temperature is
found to be 16100 $\pm$ 500 K.   As discussed in Evans, et al. (2023), errors in the temperature were estimated by visual inspection of the model comparisons to identify clear differences from the preferred solution.

\hfill\vfill
\eject

\section{Discussion/ Conclusions}

\begin{figure}
 \includegraphics[width=3.0in]{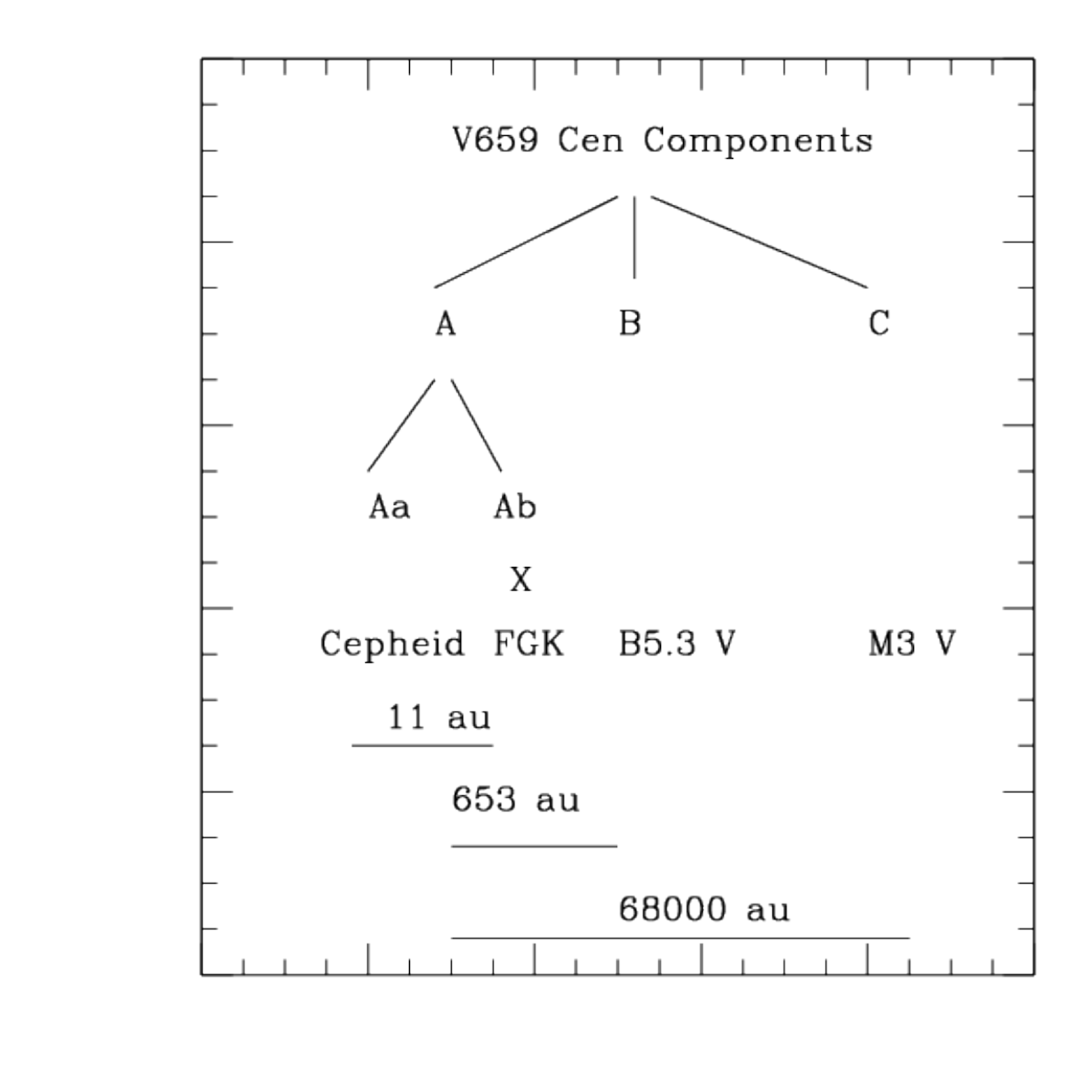}
\caption{ The schematic summary of the menbers of the V659 Cen system. The components are at the top.  Below them is the information about the spectral type.  Below that is an estimate of the separation. 
    \label{comp.sum}}
\end{figure}

Deconstructing the V659 Cen system has required ground-based observations where the Cepheid
dominates the spectrum, and also ultraviolet and X-ray observations where the hottest star
and the least massive star dominate respectively.  The new STIS spectrum shows clearly
the structure of the system:  the companion in the spectroscopic binary is the low
mass companion;  the hottest star is at a wider separation.
The system members are summarized in Table~\ref{members}  and in Fig.~\ref{comp.sum}.
 Component Ab is labeled ``FGK'' in the table because those are the main sequence spectral types which would be detected in the X-ray exposure.
A correllary of this hierarchy is
that the temperature and its related mass of the hottest star
cannot be used to determine the mass of the
Cepheid as was done for V350 Sgr (Evans, et al. 2025).  Nor would velocities from high resolution spectra
in the ultraviolet provide the mass ratio with the Cepheid from the spectroscopic
binary.  

\begin{deluxetable}{lllr}
\tablecaption{System Members\label{members}}
\tablewidth{0pt}
\tablehead{  \colhead{} &
  \colhead{} &  \colhead{ Sep }  & \colhead{Spectral}   \\
    \colhead{} &\colhead{} &  \colhead{ `` }  & \colhead{Type}   \\
}
\startdata
Aa & Cepheid  &   &   \\
Ab & Companion 1 &  0.01 & FGK \\
B & Companion 2 &  0.6 & B5.3 \\
 &   &  &  \\
C & Wide  & 62 & M3 V \\
\enddata
\end{deluxetable}


\section{Acknowledgments}

It is a pleasure to thank Pawel Mosklik for  discussion on mode identification.
 We thank the referee for improved clarity in the presentation.
We thank all the observers of the VELOCE project for their continued support.

Support for JK and CP were provided from HST-GO-15861.001-A.
Institutional support was provided to NRE by the Chandra X-ray Center NASA Contract NAS8-03060.
The research leading to these results has received funding to P. K. from the European Research Council (ERC) under the European Union's Horizon 2020 research and innovation program (project UniverScale, grant agreement 951549).
This research has been supported by the Polish-French Marie Sk{\l}odowska-Curie and Pierre Curie Science Prize awarded by the Foundation for Polish Science. P. K. acknowledges the support of the French Agence Nationale de la Recherche (ANR), under grant ANR-23-CE31-0009-01 (Unlock-pfactor).
A. G. acknowledges the support of the Agencia Nacional de Investigación Científica y Desarrollo (ANID) through the FONDECYT Regular grant 1241073.
This project has received funding from the European Research Council (ERC) under the
European Union's Horizon 2020 research and innovation programme (Grant Agreement
No. 947660). RIA is funded by the Swiss National Science Foundation through an
Eccellenza Professorial Fellowship (award PCEFP2\_194638).  The Euler telescope is
funded by the Swiss National Science Foundation (SNSF). S.S. would like to
acknowledge the Research Foundation-Flanders (grant No.: 1239522N).

This work has made
use of data from the European Space Agency (ESA) mission Gaia (https://www.cosmos.esa.int/gaia), processed by
the Gaia Data Processing and Analysis Consortium (DPAC, https://www.cosmos.esa.int/web/gaia/dpac/consortium).
Funding for the DPAC has been provided by national institutions, in particular the institutions participating in the
Gaia Multilateral Agreement.
The SIMBAD database, and NASA’s Astrophysics Data System Bibliographic Services
were used in the preparation of this paper.

The data presented in this article were obtained from the Mikulski Archive for Space Telescopes (MAST) at the Space Telescope Science Institute. The specific observations analyzed can be accessed
via \dataset[10.17909/6er1-f837]{http://dx.doi.org}.




\end{document}